# Large Predicted Self-Field Critical Current Enhancements In Superconducting Strips Using Magnetic Screens

Yu. A. Genenko,[1,2] A. Usoskin,[3] and H. C. Freyhardt[1,3]

[1]*Institut für Materialphysik, Universität Göttingen, Hospitalstrasse 3-7, D-37073, Göttingen, Germany*
[2]*Donetsk Physical and Technical Institute, National Academy of Science Ukraine, 340114 Donetsk, Ukraine*
[3]*Zentum für Funktionswerkstoffe gGmbH, Windausweg 2, D-37073, Göttingen, Germany*
(Received 16 February 1999)

A transport current distribution over a wide superconducting sheet is shown to strongly change in the presence of bulk magnetic screens of a soft magnet with a high permeability. Depending on the geometry, the effect may drastically suppress or protect the Meissner state of the sheet through the enhancement or suppression of the edge barrier critical current. The total transport current in the magnetically screened Meissner state is expected to compete with the critical current of the flux-filled sheet only for samples whose critical current is initially essentially controlled by the edge barrier effect.

Superconductivity is known to be suppressed by external magnetic fields and magnetic impurities. In this Letter it is predicted that magnetic materials used for the conditioning of a magnetic field of a current-carrying conductor can drastically increase its loss-free current.

There are two alternative loss-free current-carrying states of a superconductor: flux-free, or Meissner, state when the magnetic flux is completely expelled from the sample, and a flux-filled critical state [1] when the magnetic flux penetrates (also partly) the sample but is pinned by material inhomogeneities which prevents its motion and accompanying dissipation. The total current in the Meissner state is normally much smaller than that in the critical state and is not considered as competitive with the latter in large-current applications. In this Letter, we show that a superconductor sheet possessing a pronounced edge barrier (of any nature) against a magnetic flux entry and surrounded by specially designed magnetic shields may exhibit the Meissner state with a transport current higher than that in the critical state.

We consider a magnetostatic problem of the Meissner state of a long thin superconductor sheet carrying a current $I$ and placed between bulk soft magnets possessing a high magnetic permeability. In this geometry, the problem may be reduced to the study of one-dimensional current distribution over the sheet.

Let the sheet occupy the region $|y| \leq d/2, |x| \leq W/2$ and be infinite along the $z$ axis. Below, we use dimensionless length units so that the half-width of the sheet $W/2 = 1$. The magnetic field produced by the isolated thin sheet has no $z$ component and may be expressed, to an accuracy of $d \ll 1$, by using the Ampere law as in [2]:

$$\mathbf{H}(x,y) = \frac{1}{2\pi} \int_{-1}^{1} du\, J(u) \frac{(-y, x-u)}{(x-u)^2 + y^2}, \quad (1)$$

where the sheet current $J(x)$ is the transport current density integrated over the sheet thickness.

The Meissner state of the sheet means the vanishing of the $H_y$ component of the field at the sheet surface. This gives the condition of the virgin state

$$H_y(x,0) = \frac{1}{2\pi} \int_{-1}^{1} \frac{du\, J(u)}{x-u} \equiv 0 \quad (2)$$

for all $x \in (-1, 1)$. The solution to this equation [3],

$$J(x) = \frac{I}{\pi\sqrt{1-x^2}}, \quad (3)$$

is physically valid not too close to sheet edges, where it is violated when $|x \pm 1| < \delta = \max\{d, \Lambda\}$, where $\Lambda = \lambda^2/d$ is the typical penetration depth of films and $\lambda$ is the magnetic field penetration depth in the bulk material.

The solution (3) diverges at the sheet edges where it, in fact, saturates on the scale of $\delta$ and achieves the maximum value of $J_m = I/\pi\sqrt{2\delta}$. The virgin state of the sheet is saved unless $J_m$ exceeds some critical value $J_\Gamma$ at which flux first enters the sample from the edges.

In samples of rectangular cross section, $J_\Gamma$ is determined by the balance between the current-induced Lorentz force exerted upon magnetic vortices at the sheet edge and the geometrical barrier force [4,5]. This gives $J_\Gamma = 2H_{c1}$ which produces, perpendicular to the sheet, the field component equal to $H_{c1}$, the lower critical field of the bulk material, at the edges and an average sheet current $J_b = \pi H_{c1}\sqrt{2\delta}$.

For thin films with $d \ll \lambda$ and clear-cut edges [6] or thicker sheets with carefully rounded edges [7] the entry of vortices is determined by the Bean-Livingston barrier [8] which allows $J_\Gamma$ as large as the Ginzburg-Landau depairing current [9]. In what follows, we assume the Meissner state to be protected by the most robust geometrical barrier [4,5].

Consider now what happens with the same sheet inserted between the two magnetic half spaces with the boundaries $y = \pm a$ parallel to the sheet, as shown in Fig. 1a. The magnetic field may be presented by means of a vector potential as $\mathbf{H} = \mu_i^{-1}\,\mathrm{curl}\mathbf{A}$, [10], where $\mathbf{A} = [0, 0, A(x,y)]$

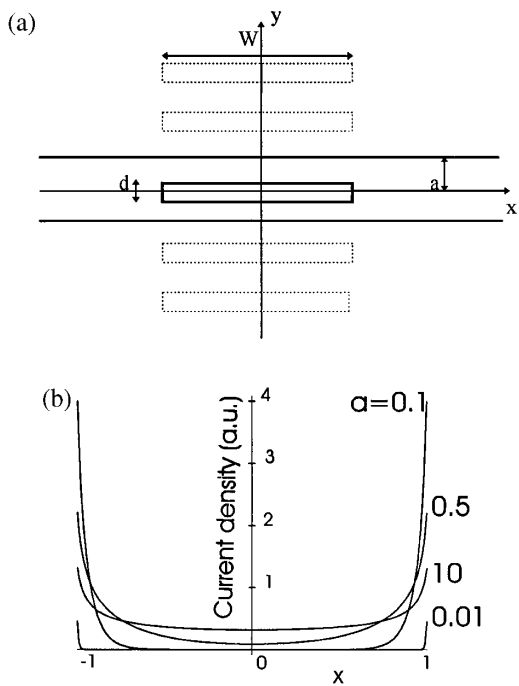

FIG. 1. (a) Current-carrying sheet parallel to boundaries of two magnet half spaces and its "images." (b) Current distribution over the sheet for various distances $a$ from the sheet to magnets (in units of the sheet half-width $W/2$).

satisfies the scalar Poisson equation

$$\Delta A = -\mu_i J(x)\delta(y), \quad (4)$$

where $\mu_i = \mu_1$ denotes the magnetic permeability in between the magnets and $\mu_i = \mu_2$ inside the magnets. The vector potential $\mathbf{A}$ and its normal to surface derivatives $\mu_i^{-1}\partial\mathbf{A}/\partial n$ should be continuous at magnet boundaries.

The solution to Eq. (4) for the arbitrary sheet current $J(x)$ may be constructed by means of the method of images [10]. The sheet produces an infinite succession of equidistant images separated by the spacing $2a$ inside the magnets, as shown in Fig. 1a. Matching the vector potential and its derivatives at boundaries $y = \pm a$, one finds a solution between the magnets,

$$A_1 = -\frac{\mu_1}{4\pi}\sum_n q^{|n|}$$
$$\times \int_{-1}^{1} du\, J(u)\ln[(y - 2an)^2 + (u - x)^2],$$

where the summation extends over all integers, and inside the magnets ($\pm y \geq a$)

$$A_2 = -\frac{\mu_1 \mu_2}{2\pi(\mu_1 + \mu_2)}\sum_{n=0}^{\infty} q^n$$
$$\times \int_{-1}^{1} du\, J(u)\ln[(y \pm 2an)^2 + (u - x)^2],$$

where a parameter $q = (\mu_2 - \mu_1)/(\mu_2 + \mu_1) < 1$.

3046

The Meissner state is then defined by the condition

$$H_y(x, 0) = \int_{-1}^{1}\frac{du}{2\pi} J(u)\sum_n q^{|n|}\frac{x - u}{(x - u)^2 + (2an)^2}$$
$$\equiv 0. \quad (5)$$

Here we consider the analytically solvable case of the soft magnetic material with a very large $\mu_2 \gg \mu_1$. In this case, $q \to 1$ and Eq. (5) converts to the equation

$$\int_{-1}^{1} du\, J(u)\coth\left(\frac{\pi}{2a}(x - u)\right) = 0. \quad (6)$$

Singular integral equations similar to Eq. (6) are known in the problem of an incompressible ideal fluid flowing through a regular cascade [11]. Let us introduce new variables $\tau = \tanh\pi u/2a$ and $t = \tanh\pi x/2a$, which vary on the interval $(-l, l)$ with $l = \tanh\pi/2a$. Substituting them into Eq. (6), one finds a divergent solution,

$$J(x) = \frac{I}{a}\frac{\cosh\pi x/2a}{\sqrt{2\cosh\pi/a - 2\cosh\pi x/a}}. \quad (7)$$

The transformation of the sheet current during the change of the distance $a$ between the sheet and the magnets is shown in Fig. 1b. At distances large compared to the sheet width, the sheet current reproduces the pattern (3) of the isolated film, but for the distances $a \ll 1$ the current distribution changes drastically. The current is pushed from the inner part of the sheet out to the edges and is exponentially small at distances larger than $a$ from the edges. The maximum sheet current at the edge is then enhanced by the factor $\sqrt{\pi/2a}$ that leads to the reduction of the total current by the same factor.

Let us recall, that the above picture of current redistribution makes sense down to $a \simeq \delta \ll 1$, where Eqs. (5) and (6) are no longer valid. In the region $a > \delta$, the maximum averaged sheet current $J_c$, by which the Meissner state is still saved, reduces with a decrease of $a$ as

$$J_c = \pi H_{c1}\sqrt{2\delta}\sqrt{\frac{\tanh\pi/2a}{\pi/2a}}. \quad (8)$$

Let us now consider the case of a transverse position of the sheet with respect to magnets as shown in Fig. 2a. The boundaries now assume locations $x = \pm a$, where $a > 1$ holds in this geometry. The images of the sheet are now centered in points $x = 2an$ with $n$ the integer. The solution to Eq. (4) in between the magnets now reads [even $J(u)$ is assumed]

$$A_1 = -\frac{\mu_1}{4\pi}\sum_n q^{|n|}$$
$$\times \int_{-1}^{1} du\, J(u)\ln[y^2 + (x - u - 2an)^2],$$

and inside the magnets ($\pm x \geq a$) it is

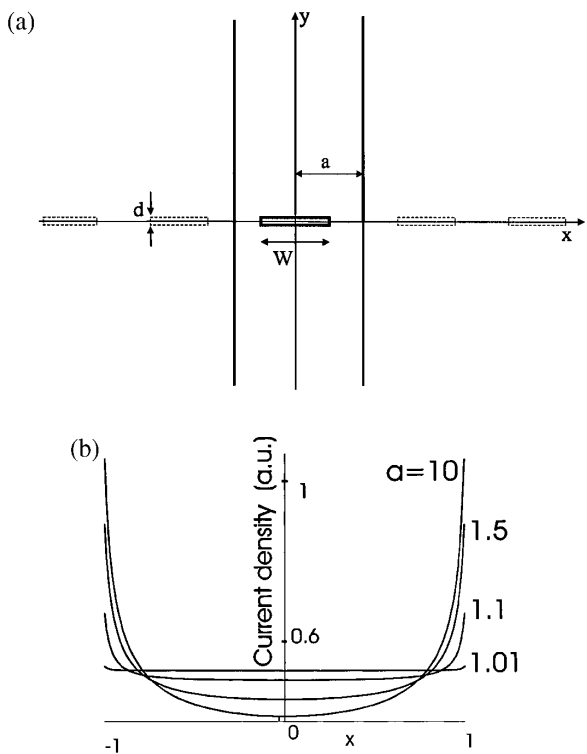

FIG. 2. (a) Current-carrying sheet perpendicular to boundaries of magnet half spaces and its images. (b) Current distribution over the sheet for various distances $a$ from the sheet center to magnets (in units of the sheet half-width $W/2$).

$$A_2 = -\frac{\mu_1\mu_2}{2\pi(\mu_1 + \mu_2)}$$
$$\times \sum_{n=0}^{\infty} q^n \int_{-1}^{1} du\, J(u) \ln[y^2 + (x - u \pm 2an)^2].$$

The condition of the Meissner state gives an equation

$$H_y(x,0) = \int_{-1}^{1} \frac{du}{4a} J(u) \cot\left(\frac{\pi}{2a}(x-u)\right) = 0. \quad (9)$$

We manage with this equation similar to Eq. (6). Using new variables $\tau = \tan\pi u/2a$, $t = \tan\pi x/2a$, which are defined within the interval $(-c, c)$, with $c = \tan\pi/2a$, we find the sheet current in this configuration,

$$J(x) = \frac{I}{a} \frac{\cos\pi x/2a}{\sqrt{2\cos\pi x/a - 2\cos\pi/a}}. \quad (10)$$

The evolution of the current distribution in the dependence on the distance $a$ to the magnets is shown in Fig. 2b. The approach of the magnets to the sheet ($a \to 1$) causes a flattening of the current distribution and a reduction of the current peaks at the edges by a factor of $\sqrt{1/(a-1)}$. This means that the enhancement of the total current provided by the edge barrier is by the same factor. By taking formally $a = 1$, one finds from Eq. (10) $J(x) = I/2 = $ const. In fact, a maximum possible current peak reduction at the edge may be achieved at the smallest reasonable distance between magnets and the sheet $a - 1 \simeq \delta$, which provides the averaged sheet current $J_c = H_{c1}\sqrt{3}$.

At distances $a - 1 > \delta$, the maximum current at which the Meissner state is still saved increases with decreasing $a$ as

$$J_c = \pi H_{c1}\sqrt{2\delta} \sqrt{\frac{\tan\pi/2a}{\pi/2a}}. \quad (11)$$

The change in the current distribution in the configurations of Figs. 1a and 2a may simply be interpreted as follows. In the first case, when $a \ll 1$, the current sheet, together with images, forms an effective current-carrying "superconducting slab" filling the space between the planes $x = -1$ and $x = 1$. The current in this "slab" tends to flow in the surface layers close to $x = \pm 1$, which is typical of bulk superconductors. In the configuration of Fig. 2a, the sheet, together with its images, forms, at $a \to 1$, an effective current-carrying "superconducting sheet" infinite in the $x$ direction. Such a "sheet" filled with a constant sheet current induces only a field parallel to the sheet which provides its flux-free state. Our results in the limit $q \to 1$ are close to those obtained in a study of a field distribution over a stack and array of superconducting films performed in Ref. [12], where equations similar to our Eqs. (6) and (9) were found.

The average Meissner current in the magnetic surrounding of Fig. 2a may be, at most, close to the value $J_\Gamma$ determined by some edge barrier. Now we show that, by a special choice of the form of magnetic banks, the current-carrying capability of a sheet in a Meissner state may be further strongly enlarged. To demonstrate this, we consider for simplicity only the case of a magnet material with $\mu_2 \gg \mu_1$ directly contacting a sheet ($a \to 1$).

At $q \to 1$, the magnetic field lines are practically perpendicular to magnet boundaries from the side of the media with smaller permeability $\mu_1$ as is true in the case of an electrostatic field near the boundary of a conductor. This enables us to use a conformal mapping of the geometry shown in Fig. 2a to find field and current distributions in the Meissner state for other magnet configurations.

Let us introduce complex variables $s = x + iy$ and $w = \zeta + i\eta$ and define an analytical function $w = [(1 + s)/2]^p - [(1 - s)/2]^p$, $p > 0$. This mapping keeps the sheet position $(-1, 1)$ on the horizontal axis $\zeta$ and bends the bulk boundaries in a variety of ways, depending on the $p$ value. The angle between the intersecting boundaries at points $\zeta = \pm 1$ equals $p\pi$. An example of the transformed system in new coordinates $(\zeta, \eta)$ is shown in Fig. 3a for $p = 1/2$.

If a vector potential $A_1$ satisfies Eq. (4) with a boundary condition $\partial A_1/\partial n = 0$ (Neumann problem) in the configuration of Fig. 2a and a corresponding magnetic field satisfies the Meissner state condition $H_y(x, 0) = 0$, then a vector potential $\Omega_1(\zeta, \eta) \equiv A_1[x(\zeta, \eta), y(\zeta, \eta)]$ satisfies Eq. (4) presented in new coordinates $(\zeta, \eta)$ and the same boundary condition on new boundaries (Fig. 3a). The Meissner state condition $H_\eta(\zeta, 0) = -\partial\Omega_1/\partial\zeta(\eta = 0) = 0$ also holds for all $\zeta$ within the interval $(-1, 1)$.

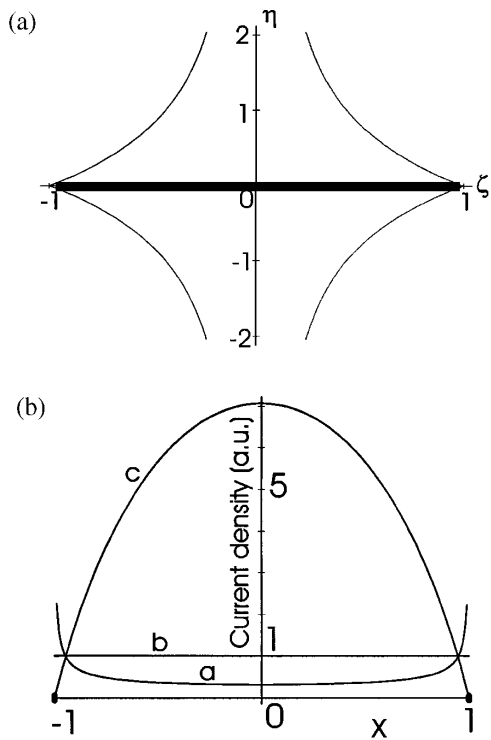

FIG. 3. (a) The current sheet in an open magnetic cavity obtained by a conformal mapping of the configuration in Fig. 2a. (b) Current distributions over the flux-free sheet. The curves $a$, $b$, and $c$ correspond to the isolated sheet, the sheet perpendicular to magnets (Fig. 2a), and the sheet in the convex cavity (Fig. 3a), respectively. $W/d = 10$.

The current distribution over the sheet is given by the jump of the parallel field component at the sheet $H_\zeta = \partial \Omega_1/\partial \eta$. For $q = a = 1$ and $p = 1/2$, one finds

$$J(\zeta) = H_\zeta(\zeta, \eta)|_{\eta=+0}^{\eta=-0} = I \frac{2(1 - \zeta^2)}{\sqrt{2 - \zeta^2}}. \quad (12)$$

The sheet current vanishes linearly at the sheet edges (generally, it vanishes as $|\zeta \pm 1|^{1/p-1}$ for $p < 1$) which enlarges extremely the current-carrying capability of the sheet.

The Meissner state holds until the sheet current at the edge ($\zeta \simeq 1 - \delta$) achieves $J_\Gamma$. The filling of the sheet with a transport current is shown in Fig. 3b for various magnet geometries. In the configuration of Fig. 3a, the average sheet current achieves a magnitude of $H_{c1}/4\delta$.

Convenient for the experimental proof of the theory would be low-pinning single crystals of BSCCO [4,5] or YBCO [13] whose critical currents are controlled by the edge barrier or very thin YBCO films whose edge barrier current $J_\Gamma$ is expected to exceed the bulk pinning critical current. For a YBCO epitaxial film of thickness $d = 50$ nm and width $W = 5$ mm with $\lambda = 0.5$ $\mu$m [14] and $\xi = 3.6$ nm [15] at 77 K, one finds $H_{c1} \simeq 2.6 \times 10^3$ A/m and critical current density at the edge $j_\Gamma = J_\Gamma/d = 10^7$ A/cm$^2$. Then, for the gap of 0.1 mm between the magnet and the sheet ($a = 1.04$), Eq. (11) gives $j_c = J_c/d = 3.3 \times 10^6$ A/cm$^2$, where a bulk pin-ning critical current typical of wide YBCO films may prevail. A total current of the film in the open convex cavity (Fig. 3a) with the same distance to magnets should be 1 order larger.

High Meissner currents in the geometries of Figs. 2a and 3a may enable vortex penetration from the top and bottom film surfaces that strongly restricts currents in anisotropic sheets possessing low parallel critical fields $H_{c1}^\parallel$. However, it is not the case for thin films with $d \ll \lambda$, where $H_{c1}^\parallel$ should be substituted by $H_{c1}(d) = (2\Phi_0/\pi d^2 \Gamma)\ln(2\Gamma d/\pi \xi)$ [16], where anisotropy parameters $\Gamma \simeq 6$ and $\Gamma \simeq 60$ are known for YBCO and BSCCO, respectively. The maximum physically possible depairing current [9] would produce on surfaces a field smaller than $H_{c1}(d)$ at $d < [12\sqrt{3}\,\lambda^2 \xi \ln(2\Gamma d/\xi)/\Gamma \ln(W/d)]^{1/3}$ which gives, for both YBCO and BSCCO, the upper estimate of thickness $d_c \simeq 0.1$ $\mu$m.

In conclusion, we have derived analytical solutions for a current distribution over the flux-free superconductor sheet surrounded by a soft high permeability magnet of various configurations. In special geometries, thin sheets possessing a pronounced edge barrier are shown to possibly exhibit transport currents exceeding bulk pinning currents.

Yu. A. G. is grateful to the Alexander von Humboldt Foundation for financial support.

[1] A. M. Campbell and J. E. Evetts, *Critical Currents in Superconductors* (Taylor & Francis, London, 1972).
[2] E. H. Brandt and M. Indenbom, Phys. Rev. B **48**, 12 893 (1993).
[3] E. H. Rhoderick and E. M. Wilson, Nature (London) **194**, 1167 (1962); A. I. Larkin, and Yu. N. Ovchinnikov, Sov. Phys. JETP **34**, 651 (1972).
[4] M. V. Indenbom, H. Kronmüller, T. W. Li, P. H. Kes, and A. A. Menovsky, Physica (Amsterdam) **222C**, 203 (1994).
[5] E. Zeldov, A. I. Larkin, V. B. Geshkenbein, M. Konczykowski, D. Majer, B. Khajkovich, V. M. Vinokur, and H. Shtrikman, Phys. Rev. Lett. **73**, 1428 (1994).
[6] K. K. Likharev, Izv. VUZOV Radiofiz. **14**, 909 (1971).
[7] M. Benkraouda and J. R. Clem, Phys. Rev. B **58**, 15 103 (1998).
[8] C. P. Bean and J. D. Livingston, Phys. Rev. Lett. **12**, 14 (1964).
[9] J. Bardeen, Rev. Mod. Phys. **34**, 667 (1962).
[10] D. Jackson, *Classical Electrodynamics* (Wiley, New York, 1962).
[11] E. Meister, Arch. Ration. Anal. **6**, 198 (1960).
[12] Y. Mawatari, Phys. Rev. B **54**, 13 215 (1996).
[13] M. W. Gardner, S. A. Govorkov, R. Liang, D. A. Bonn, J. F. Carolan, and H. W. Hardy, J. Appl. Phys. **83**, 3714 (1998).
[14] A. T. Fiory, A. F. Hebard, P. M. Mankievich, and R. E. Howard, Appl. Phys. Lett. **52**, 2165 (1988).
[15] M. N. Kunchur, D. K. Christen, and J. M. Phyllpis, Phys. Rev. Lett. **70**, 998 (1993).
[16] A. A. Abrikosov, Sov. Phys. JETP **19**, 988 (1964).